==================

\documentstyle[12pt]{article}

\newcommand{\be}{\begin{equation}}
\newcommand{\ee}{\end{equation}}
\newcommand{\bm}{{\bf M}}

\newcommand{\bx}{{\bf x}}

\newcommand{\sech}{{\rm sech}}
\newcommand{\qplus}{\mbox{$q+Q_{0}$}}
\newcommand{\qmin}{\mbox{$q-Q_{0}$}}
\newcommand{\xplus}{\mbox{$x+x_{0}$}}
\newcommand{\xmin}{\mbox{$x-x_{0}$}}
\newcommand{\sechsp}{\mbox{${\sech}^{2}({\xplus})$}}
\newcommand{\sechsm}{\mbox{${\sech}^{2}({\xmin})$}}

\begin{document}

\begin{titlepage}

\begin{flushright}
UBC TP-93-19 \\
October, 1993
\end{flushright}

\vspace{0.4in}

\begin{center}
{\Large On the Observability of Meso- or Macro-scopic
	Quantum Coherence
	 of Domain Walls
	 in Magnetic Insulators}  \\
\vspace{0.4in}
{\large Frank Gaitan} \\
\vspace{0.35in}
{\em Department of Physics} \\
{\em University of British Columbia} \\
{\em 6224 Agriculture Road} \\
{\em Vancouver, British Columbia, CANADA, V6T 1Z1} \\
\vspace{0.5in}
{\bf ABSTRACT}
\end{center}

\vspace{0.15in}

\noindent Results are presented of a numerical calculation of the tunneling
gap for a domain wall moving in the double well potential of a pair of voids
in a magnetic insulator. Both symmetric and asymmetric double well potentials
are considered. It is found that, even in the absence of dissipation, the
prospect for observing quantum coherence on a meso- or macro-scopic scale
appears unlikely.

\vspace{0.3in}

\begin{flushright}
PACS Indices: 75.60-d; 75.60Ch \\
              75.60.Lr
\end{flushright}

\end{titlepage}

\pagebreak

There has been a great deal of interest recently in the prospect that
magnetic systems might provide a new setting in which to observe a
macroscopic degree of freedom behaving quantum mechanically \cite{stamp1}.
To date, the magnetic systems considered are: (i) magnetic grains
\cite{c+g,euro,b+c,awsch1,awsch2}; and (ii) solitons in magnetic systems
(particularly $180^{\circ}$-Bloch walls)
\cite{stamp1,stamp2,ci+s,gio,exp1,exp2}.
Attention
has focused on macroscopic quantum tunneling since the conditions necessary for
its observation appear the most favorable \cite{c+l}. Arguably, of more
fundamental interest is macroscopic quantum coherence (MQC) because of its
connections with quantum measurement theory \cite{legg}. In MQC the macroscopic
object tunnels {\em periodically}
through the central barrier of a double well
potential (DWP). This effect is a direct consequence of the
quantum state of the object
being in a coherent linear superposition of macroscopically distinguishable
states, and it is this aspect of MQC that connects it to the Schrodinger
cat paradox \cite{schro} and to quantum measurement theory.

In this paper we will examine quantum coherence (QC) for a $180^{\circ}$-Bloch
wall (we suppress the ``$180^{\circ}$'' below) moving
in the double well potential
due to a pair of voids present in a uniaxial magnetic insulator with
quality factor $Q=K/2\pi M^{2}\gg 1$ ($K$ =
magnetic anisotropy constant,
$M$ = spontaneous magnetization) at $T=0$ \cite{mal}. Under these conditions:
(i) demagnetization effects are small so that the voids do not alter
significantly the Bloch wall configuration of the magnetization \cite{s&g};
and (ii) most dissipative effects are small \cite{ci+s}.
Thus our system,
under the conditions assumed, is expected to have weak dissipation---a
situation we approximate by ignoring dissipation (though see below).
We calculate the ground (first excited) state energy $E_{0}$
($E_{1}$) numerically (from which the tunneling gap is $\Delta_{0}=E_{1}-
E_{0}$) for: (i) identical voids leading to a symmetric double well potential
for various wall sizes $N$ (the number of spins in the wall) and void
separations $L$; and (ii) non-identical voids leading to an asymmetric
double well potential for varying degrees of asymmetry (for a
particular choice of $N$ and $L$). We find that observation of quantum
coherence on either a macro- or meso-scopic scale appears unlikely. For
macroscopic QC ($N\ge 10^{4}$), weak stray magnetic fields introduce a bias
into the gap which masks $\Delta_{0}$ except for void separations very close
to the value at which the central barrier disappears. At these
separations, the tunneling gap varies on a length scale that is less
than the coarse graining length scale. One would expect that the
experimentally relevant
gap would be a coarse grained average which is seen to be
less than the bias and so unobservable. For mesoscopic
QC ($N\sim 10^{2}-10^{3}$), although the bias introduced by a stray magnetic
field is quite small, the slightest bit of asymmetry in the voids is
sufficient to pin the wall to the larger of the two voids. The difficulty here
is that an adequately large tunneling gap requires a very small central barrier
which is destroyed by the slightest difference in the voids.
For spherical
voids, we find that pinning of the wall
can be avoided only when the difference
in the radii is much less than the coarse graining length scale.
Averaging the effects of asymmetry over the coarse graining length scale
leads one to conclude that a real mesoscopic wall will likely be pinned
by asymmetry. Thus even in the dissipationless
approximation, one expects that observation of quantum coherence on
any scale larger than
microscopic appears unlikely due to the severe tolerances imposed on the
experimental situation.

The system of interest is a magnetic insulator which is a lattice of spins
(with lattice constant $l_{0}$) coupled to each other via the exchange
and dipole-dipole interactions; and to the underlying lattice via the
anisotropy interaction which is assumed to be uniaxial with easy axis
along $\hat{z}$.
For the length scales of interest to us the lattice system
can be coarse grained so that the magnetic state of the system is described
by a magnetization ${\bf M}({\bf x},t)$ defined on a 3-D spatial continuum.
The total static energy in the absence of
an external magnetic field is the sum of the exchange, anisotropy and
demagnetization energies. A stationary Bloch wall is a soliton
configuration of the
magnetization \bm(\bx) with vanishing demagnetization energy, subject to the
boundary condition $\bm(y\rightarrow\pm\infty)=\pm M\hat{z}$. The spatial
variation of the wall is localized to a planar region of thickness $\lambda =
\sqrt{J/K}$ ($J$ = exchange stiffness constant) which is assumed to be parallel
to the xz-plane. The wall coordinate $q$ specifies the distance
from the origin to a reference
point on the wall (viz.\ $\bx_{wall}=q\hat{y}$) \cite{mal}.

Voids in the magnetic insulator act as pinning sites for the wall. For
materials
with $Q\gg 1$, the attraction is due primarily to a reduction in the exchange
and anisotropy energies that occurs when the wall sits on the void. For a
void of length scale $R$ satisfying $l_{0}\ll R\ll\lambda$ ($l_{0}$ = lattice
constant $\equiv 5\AA$), located at the origin, the pinning potential seen by
a flat Bloch wall is $U(q)=-U_{0}\,{\rm sech}^{2}(q/\lambda)$ \cite{alt}, where
$U_{0}=2KV_{d}$
and $V_{d}$ is the void volume. In our calculation, $\lambda
=1000\AA$ ($50\AA$) for walls with $N\ge 10^{4}$ ($300\le N \le 3000$)
and $U_{0}=0.1{\rm eV}$ ($1.0\times 10^{-5}{\rm eV}$).
For LaGaYIG, $Q=25.2$ and $K\sim 2000
\,{\rm ergs}/{\rm cm}^{3}$, so that for a spherical void $R\sim 200\AA$
($10\AA$).
We consider two spherical voids located at $\bx_{\pm}=\pm Q_{0}\hat{y}$
($L=2Q_{0}$) with volumes $V_{+}=aV_{-}$ ($a\ge 1$). They produce the double
well pinning potential
\begin{displaymath}
U(q)=-U_{0}\left[\,\sech^{2} \left(\frac{\qplus}{\lambda}\right)+a\,\sech^{2}
      \left(\frac
       {\qmin}{\lambda}\right)\,\right] .
\end{displaymath}
When $a=1$ we obtain a symmetric double well potential (SDWP); otherwise, the
wall sees an asymmetric double well potential (AsDWP). For an energy $E$
corresponding to QC there will be 4 turning points $T_{1}<T_{2}<T_{3}<
T_{4}$. We refer to the region $q<T_{1}$ as the ``left barrier''; the
region $T_{1}<q<T_{2}$ as the ``left well''; the region $T_{2}<q<T_{3}$ as
the ``central barrier''; the region $T_{3}<q<T_{4}$ as the ``right well''; and
the region $T_{4}<q$ as the ``right barrier''. Varying the void separation
$L$ (viz.~$Q_{0}$) varies the depth of the wells and the height and width of
the central barrier. The experimental situation envisioned is either: (1) a
thin
film or; (2) a very narrow wire of the magnetic insulator
in which only one Bloch wall is present \cite{ci+s}. In the thin film case,
only a small region (of the wall) of cross-sectional area $A_{w}$ is
involved in tunneling between the pinning sites \cite{ci+s}.
In our analysis below, we treat the wall as if it were flat, whereas,
for the thin film scenario, it will in fact
be curved.  Curvature effects will be discussed below (see also \cite{s&g}).
In the case of the very thin wire,
curvature effects are not expected  to be important because of
the large energy required to bend the wall on a length scale of order
the cross-sectional dimension of the wire.
Flat moving walls have a kinetic energy $M_{D}\dot{q}^{2}/2$ \cite{mal}. Here
$M_{D}=A_{w}/2\pi\gamma^{2}\lambda$ is the D\"{o}ring mass;
and $\gamma$ is the gyromagnetic ratio. The wall
Hamiltonian is $H=p^{2}/2M_{D}+U(q)$. Introducing the dimensionless length $x=
q/\lambda$ and the energy scale ${\cal S}=\hbar^{2}/2M_{D}\lambda^{2}$, we
can write the time independent Schrodinger equation in the dimensionless form
\begin{displaymath}
\left[\,\frac{d^{2}}{dx^{2}}+{\cal U}_{0}\left\{\,\sechsp +a\,\sechsm\,\right\}
    +{\cal E}\,\right]\psi = 0  .
\end{displaymath}
Here ${\cal U}_{0}=U_{0}/{\cal S}$; $x_{0}=Q_{0}/\lambda$; and ${\cal E}=E/
{\cal S}$. The dimensionless potential strength ${\cal U}_{0}$ is related to
the wall size $N=A_{w}\lambda/l_{0}^{3}$ by
\begin{displaymath}
{\cal U}_{0}=\left(\frac{U_{0}l_{0}^{3}}{\pi^{2}g^{2}\mu^{2}_{B}}\right)N,
\end{displaymath}
where $g$ is the electron $g$-factor; and $\mu_{B}$ is the Bohr magneton.

We utilize a shooting algorithm \cite{numrec} to solve the eigenvalue
problem numerically. Details of this calculation will be given in
\cite{s&g}.
Our results for the tunneling gap $\Delta_{0}$
appear in Tables 1 and
2 and correspond to macroscopic and mesoscopic
walls of thickness $\lambda =1000\AA$ and $50\AA$
respectively.
So far we have assumed our system of
wall and voids to be completely isolated.
If a weak stray magnetic field ${\bf H}_{ext}$
were present, it would produce a
bias $\epsilon$ in the gap $\Delta = \sqrt{\Delta_{0}^{2}+\epsilon^{2}}$. This
bias is a consequence of the Zeeman energy density $-\bm\cdot{\bf H}_{ext}$.
For an actual
stray field, the direction of ${\bf H}_{ext}$ is unknown and the
experimentally relevant bias $\bar{\epsilon}$
is obtained by averaging over this direction. It is easily shown that
$\bar{\epsilon} \sim MA_{w}LH_{ext}$. We take $H_{ext}\sim10^{-6}G$ as
indicative of the magnitude of a stray magnetic field (AC-magnetic fields of
$10^{-5}G$ have been used in measurements of the frequency dependent
magnetic susceptibility \cite{awsch2}). Clearly, a necessary condition for
observable domain wall QC is $\Delta_{0}>\bar{\epsilon}$.
For $N\ge 10^{4}$, we find that $A_{w}=
(1.1\times 10^{-17}{\rm cm}^{2})N$ (and
$M_{D}=(5.6\times 10^{-28}{\rm gm})
N$),
so that for LaGaYIG for which $M\sim 10G$,
$\bar{\epsilon}=(1.04\times 10^{-11}K)N$. Thus the bias grows with wall size
$N$, making observation of QC more difficult for the larger walls. To proceed
further, note that there exists a limited range of void separations for
which $\Delta_{0}$ corresponds to QC {\em and} still satisfies $\Delta_{0}>
\bar{\epsilon}$. For $L<L_{min}$, the ground state energy is above the
central barrier; while if $L>L_{max}$, $\Delta_{0}<\bar{\epsilon}$. Let
${\cal R}=L_{max}-L_{min}$ be the size of the allowed range of void
separations. Its value is given in Table 1 and is obtained (for a given $N$)
by comparing $\Delta_{0}$ and $L$ with $\bar{\epsilon}$.
Quantum coherence will be observable only when: (i) the uncertainty in
$L$ satisfies
$\Delta L\ll{\cal R}$; and (ii) ${\cal R}\gg {\cal C}$, where
${\cal C}\sim (2-3)l_{0}$ is the coarse graining length scale.
One expects that $\Delta L\sim l_{0}$ and in our
calculation $l_{0}=5\AA$. If either (or both) of these conditions is (are)
not satisfied one would expect that the experimentally relevant gap would be
an average of $\Delta_{0}$ over the appropriate length scale.
{}From Table 1 we see that, for $N\ge 10^{4}$, such an average is
necessary and that any reasonable procedure gives $\bar{\Delta}<
\bar{\epsilon}$. Thus macroscopic QC ($N\ge 10^{4}$) is not expected to be
observable due to the rapid variation of the
tunneling gap to small changes in $L$ and the large bias $\bar{\epsilon}$
introduced by a stray magnetic field. We also see that in the case of the SDWP
(in the absence of dissipation),
the conditions for observable QC do not rule out walls with
$N\sim 10^{2}-10^{3}$
(see Table 2) which would correspond to mesoscopic quantum coherence.
(Here $A_{w}=(2.67\times 10^{-16}{\rm cm}^{2})N$; $M_{D}=(2.73\times 10^{-25}
{\rm gm})N$; and $\bar{\epsilon}$ is given in Table 2.)
We now go on to examine the effects of asymmetry on the case of mesoscopic
QC.

Since $U_{0}=2KV_{d}$,
the larger the void, the more strongly it attracts the
wall. Thus, if asymmetry is sufficiently pronounced, QC is lost because the
larger void pins the wall. This effect can be seen by following the ground
state energy as we increase $a$ from 1 (see Table 3).
Imagine $a=1$ (corresponding to a
SDWP) and that $\{N,U_{0},L\}$ are such that,
in the absence of dissipation,
we have QC. Imagine further that we increase $a$ so that the void at $q=Q_{0}$
attracts the wall more strongly than the other void. This stronger attraction
causes $E_{0}$ to decrease (i.\ e.\ become more negative) as the probability
distribution in the ground state
begins to shift towards $q=Q_{0}$. As we continue
to increase $a$, we reach a critical value
$a_{\ast}$ at which $E_{0}$ is equal to the value of the AsDWP
at the metastable minimum of the left well $U_{meta}$.
For $a>a_{\ast}$, $E_{0}$ drops below the
metastable minimum which corresponds to the pinning of the wall at $q=Q_{0}$
and the destruction of QC. Intuitively, we expect that when $\Delta U_{0}=
U_{0}^{+}-U_{0}^{-}=(a-1)U_{0}^{-}$ is approximately equal to the barrier
height $U_{bh}$ of the SDWP, the larger defect will pin the wall.
For mesoscopic walls, $U_{bh}$ is given in Table 2.
For $N=300$, $L=75\AA$; $U_{bh}=2.8
\times 10^{-7}{\rm eV}$ (note the difference in units relative to Table 2).
Thus $\Delta U_{0}=U_{bh}$ corresponds to
$\overline{a_{\ast}}\approx 1.008$.
A numerical calculation of $E_{0}$ for this case gives $a_{\ast}=1.038>
\overline{a_{\ast}}$ (see Table 3).
For the spherical voids
we have
been considering, if $R_{-}=10\AA$, then $R_{+}={a_{\ast}}^{1/3}R_{-}=
10.1\AA$.
Thus if asymmetry is not to destroy QC, the radii of the two voids must
satisfy $\Delta R=R_{+}-R_{-}<0.1\AA$.
Such a tolerance is clearly unattainable and since $\Delta R\ll{\cal C}$
we must average the effects
of asymmetry over the coarse-graining length scale ${\cal C}\sim 10-15\AA$.
As the majority of $\Delta R$ values entering into the average correspond to
pinning of the wall, we conclude that asymmetry in the voids acts to destroy
QC in this case.  We might hope to
overcome this difficulty by increasing $L$
and so increasing $U_{bh}$.
For $L=103\AA$, $N=300$, the SDWP tunneling gap is $\Delta_{0}=
7.7\times 10^{-9}K$ (see Table 2). At this separation,
QC is marginally observable in the absence
of dissipation. In this case
$U_{bh}=2.8\times 10^{-6}{\rm eV}$ (see Table 2).
Then $\Delta U_{0}=U_{bh}$ gives
$\overline{a_{\ast}}=1.085$.
We did not determine $a_{\ast}$ numerically for this case. In the previous
example we saw that $(a_{\ast}-1)\sim 5(\overline{a_{\ast}}-1)$ so that
we will estimate $(a_{\ast}-1)\sim 10(\overline{a_{\ast}}-1)$ for this case.
This gives $a_{\ast}\sim 1.85$. For $R_{-}=10\AA$, $\Delta R\sim 2.3\AA$.
Again $\Delta R<{\cal C}$ so that
an average of the effects of asymmetry over ${\cal C}\sim 10-15\AA$ is
necessary. As in the previous case, the majority of the $\Delta R$ values
correspond to pinning of the wall so that we again conclude that asymmetry in
the voids will act to destroy QC in this already marginal case.
Larger values of $L$ lead to $\Delta_{0}<\bar{\epsilon}$.
We see that
asymmetry of the voids will be sufficient to destroy any remaining
vestige of domain wall QC---even in the absence of dissipation. The basic
difficulty is that maximizing the tunneling gap requires a very small
central barrier; so small in fact,
that the most minute asymmetry in the two
voids produces a bias in the double well which is of order of the height of
the central barrier and so capable of pinning the wall
(viz.\ destroying QC).
We suspect that this is
generally true of macroscopic QC: large objects require small barriers
which are easily removed by small imperfections in the experimental set-up.

In this paper we have carried out a numerical analysis of domain wall
quantum coherence in a uniaxial magnetic insulator with quality factor $Q\gg 1$
at $T=0$. We find that QC on any scale larger than microscopic appears
unlikely due to the combined effects of stray magnetic fields and asymmetry in
the voids which are responsible for producing the double well potential seen by
the domain wall. Our calculation assumed a flat wall although
curved walls are expected in the thin film scenario. For this scenario, and for
voids of given size, tunneling will only occur if $L<L_{crit}$
when curvature effects are included \cite{ci+s}. This is because
the curvature energy acts to raise the minima of the DWP relative to the top
of the central barrier, thus reducing $U_{bh}$.
When $L=L_{crit}$, the central barrier has disappeared
and we are no longer in the QC regime. If $L_{crit}>L_{max}$, the tunneling
gap becomes unobservable before curvature effects become significant.
Otherwise, $L_{crit}<L_{max}$ and curvature effects act to reduce the range of
void separation ${\cal R}\rightarrow {\cal R}_{new}=L_{crit}-L_{min}$
corresponding to QC.
For macro-walls, ${\cal R}$ was already small enough to rule out macro-QC so
that decreasing ${\cal R}\rightarrow {\cal R}_{new}$ acts to strengthen this
conclusion. For meso-walls, if ${\cal R}_{new}<{\cal C}$, then curvature
effects
have reduced the range of QC sufficiently that stray magnetic fields are
expected to make meso-QC unobservable. Finally, if ${\cal R}_{new}>{\cal C}$,
asymmetry in the void sizes can more easily destroy meso-QC since
curvature effects act to reduce $U_{bh}$.
Thus curvature effects are not expected to modify
our conclusion that observation of meso- or macro-QC of domain walls appears
unlikely. Our calculation ignores dissipation; a proper inclusion of its
effects will also act to strengthen our conclusion.
Although we have considered a particular type of defect, we expect our
conclusion to also apply when: (1) the defects are such that the pinning
potential is primarily due to a reduction in the exchange and anisotropy
energies of the wall (when pinned to the defect); and (2) the wall
configuration
${\bf M}(y)$ depends only on the wall normal coordinate $y$. Under these
conditions we continue to expect $U(q)\sim (J/\lambda^{2}+K)V_{d}{\rm sech}^{2}
(q/\lambda)$ and our results to be indicative of the scale of QC to be
expected. It should be noted that the phase coherences necessary for
establishing quantum coherence
are much more delicate than those necessary for establishing
quantum tunneling so
that our work does not preclude a priori the possibility of observing
macroscopic quantum tunneling of domain walls in magnetic insulators.
I would like to thank Philip Stamp for useful discussions, T. Howell III for
support, and NSERC of Canada for financial support.

\pagebreak

\begin{center}
   {\bf List of Table Captions}
\end{center}

\vspace{0.5in}

\begin{enumerate}
  \item Tunneling gap $\Delta_{0}$; width ${\cal R}$ of range of
        observable void separations; central barrier height $U_{bh}$;
        and bias $\bar{\epsilon}$ for
        a macroscopic Bloch wall moving in a SDWP.

  \item Tunneling gap $\Delta_{0}$; width ${\cal R}$ of range of
        observable void separations; central barrier height $U_{bh}$;
        and bias $\bar{\epsilon}$ for
        a mesoscopic Bloch wall moving in a SDWP.

  \item Asymmetry parameter $a$; ground state energy $E_{0}$;
        metastable minimum $U_{meta}$ of AsDWP; and barrier height
        $U_{bh}$ for a mesoscopic Bloch wall with $N=300$ and
        void separation $L=75\AA$.
\end{enumerate}

\pagebreak

\begin{table}[t]
\begin{tabular}{llllll} \hline
N(spins)  & $L(\AA)$ & $\Delta_{0}(K)$ & ${\cal R}(\AA)$ & $U_{bh}(K)$ &
			 $\bar{\epsilon}\,$(K)  \\ \hline
$10^{4}$ & 1320      & AB                  & $\sim 6$  & --- &
				   $1.04\times 10^{-7}$ \\
         & 1322      & $4.0\times 10^{-4}$ &  & $1.8\times 10^{-2}$ &   \\
	 & 1324      & $2.0\times 10^{-5}$ &  & $3.6\times 10^{-2}$ &   \\
	 & 1326      & $5.0\times 10^{-7}$ &  & $5.9\times 10^{-2}$ &
				       \\ \hline
$10^{5}$ & 1318      & AB                  & $\sim 2$  & --- &
				  $1.04\times 10^{-6}$ \\
         & 1319      & $1.8\times 10^{-4}$ &    & $3.0\times 10^{-3}$ &  \\
	 & 1320      & $9.6\times 10^{-6}$ &    & $6.7\times 10^{-3}$ &  \\
	 & 1321      & $2.0\times 10^{-7}$ &    & $1.2\times 10^{-2}$ &
				       \\ \hline
$10^{6}$ & 1317      & AB                  & $\sim 1$ & --- &
				  $1.04\times 10^{-5}$ \\
         & 1318      & $2.4\times 10^{-5}$ &   & $7.9\times 10^{-4}$ &  \\
	 & 1319      & $9.1\times 10^{-9}$ &   & $3.0\times 10^{-3}$ &  \\
	 & 1320      & $< 7.2\times 10^{-11}$ & & $6.7\times 10^{-3}$&
				       \\ \hline
$10^{7}$ & 1317.2    & AB                  & $<0.1$& --- &
				  $1.04\times 10^{-4}$ \\
         & 1317.3    & $2.0\times 10^{-5}$ & & $8.5\times 10^{-5}$ &  \\
	 & 1317.4    & $8.2\times 10^{-6}$ & & $1.4\times 10^{-4}$ &  \\
	 & 1317.6    & $4.0\times 10^{-7}$ & & $3.0\times 10^{-4}$ &  \\
	 & 1318.0    & $2.2\times 10^{-10}$ & & $7.9\times 10^{-4}$&
				       \\ \hline
\end{tabular}
$^{*}{\rm AB}={\rm ground\; state\; Above\; Barrier}$

\vspace{1in}

\caption{}
\end{table}

\pagebreak

\begin{table}[t]
\begin{tabular}{llllll} \hline
N(spins)  & $L(\AA)$ & $\Delta_{0}(K)$ & ${\cal R}(\AA)$ & $U_{bh}(K)$ &
			 $\bar{\epsilon}\,$(K)  \\ \hline
300    & 74      &${\rm AB}^{*}$         & $\sim 30$ & --- &
				   $5.2\times 10^{-9}$ \\
       & 75  & $1.1\times 10^{-3}$ & & $3.3\times 10^{-3}$ &   \\
         & 80      & $2.5\times 10^{-4}$ &  & $7.2\times 10^{-3}$ &       \\
	 & 103     & $7.7\times 10^{-9}$ &  & $3.2\times 10^{-2}$ &       \\
         & 110      & $2.4\times 10^{-10}$ &  & $4.0\times 10^{-2}$ &   \\
         & 120      & $1.8\times 10^{-12}$ &  & $5.2\times 10^{-2}$  &   \\
				       \\ \hline
3000 & 69      & AB                  & $\sim 10$ & --- &
				   $4.3\times 10^{-8}$ \\
         & 75      & $3.1\times 10^{-6}$ &    & $3.3\times 10^{-3}$ &  \\
	 & 80      & $6.1\times 10^{-9}$ &    & $7.2\times 10^{-3}$ &  \\
				       \\ \hline
\end{tabular}
$^{*}{\rm AB}={\rm ground\; state\; Above\; Barrier}$

\vspace{1in}

\caption{}
\end{table}

\pagebreak

\begin {table}[t]
\begin{tabular}{llll} \hline
$a$ & $E_{0}\; (K)$ & $U_{meta}\; (K)$ & $U_{bh}\; (K)$ \\ \hline
1.000 & $-1.4638\times 10^{-1}$ & $-1.4859\times 10^{-1}$ & $3.33\times
							    10^{-3}$ \\
							    \\ \hline
1.010 & $-1.4719\times 10^{-1}$ & $-1.4894\times 10^{-1}$ & $2.96\times
							    10^{-3}$ \\
							    \\ \hline
1.020 & $-1.4814\times 10^{-1}$ & $-1.4930\times 10^{-1}$ & $2.59\times
							    10^{-3}$ \\
							    \\ \hline
1.030 & $-1.4914\times 10^{-1}$ & $-1.4966\times 10^{-1}$ & $2.23\times
							    10^{-3}$ \\
							    \\ \hline
1.038 & $-1.4997\times 10^{-1}$ & $-1.4995\times 10^{-1}$ & $1.94\times
                                                         10^{-3}$ \\
							 \\ \hline
\end{tabular}

\vspace{1in}

\caption{}
\end{table}

\begin{thebibliography}{99}
\bibitem{alt} See ref.~\cite{s&g}. Note that this
	      reference corrects the expression for $U_{0}$ given in
	      ref.~\cite{stamp1}.
\end{thebibliography}
\end{document}